\documentclass{jps-cp}
\usepackage{txfonts} 

\title{Longitudinal Spin Transfer to $\Lambda$ Hyperons in CLAS12}

\author{Matthew \textsc{McEneaney}$^{1}$ for the CLAS Collaboration}

\inst{$^{1}$Duke University, Durham, NC 27708, USA}

\email{matthew.mceneaney@duke.edu}

\recdate{January 12, 2021}

\abst{Using the self-analyzing decay of the $\Lambda$, the longitudinal spin transfer $D_{LL'}$ to the hyperon from a polarized electron beam scattering off an unpolarized proton target can be determined. For $\Lambda$s produced in the current fragmentation region, this quantity is proportional to the helicity dependent fragmentation function $G_1^\Lambda$ and can provide insight into the spin structure of the $\Lambda$. Currently, limited experimental data on $D_{LL'}$ cannot discriminate between different models of $\Lambda$ spin structure. This contribution reports on the status of the ongoing analysis of the longitudinal spin transfer using data taken by the CLAS12 spectrometer at Jefferson Lab with a 10.6 GeV polarized electron beam.  We also report on the novel use of Graph Neural Networks (GNNs) to identify signal events.}

\kword{spin transfer, longitudinal polarization, hyperon, graph neural networks, clas12}

\begin{document}
\maketitle


\section{Motivation} \label{Motivation}
The decay of the $\Lambda$ hyperon to a proton pion pair has a high branching ratio of $63.9\%$ and the $\Lambda$ polarization is easily accessible from the angular dependence of the decay protons in the $\Lambda$ rest frame.  For $\Lambda$s produced in the current fragmentation region, the spin transfer coefficient $D_{LL'}$ is proportional to the helicity dependent fragmentation function $G_1^\Lambda$ \cite{Airapetian_2006}.  If $D_{LL',f}$ is the partial spin transfer from a quark of flavor $f$ to the $\Lambda$, then the spin transfer coefficient is described by the ratio between helicity dependent and helicity independent fragmentation functions $G_{1,f}^{\Lambda}$ and $D_{1,f}^{\Lambda}$ \cite{Airapetian_2006}.  
%
%
In Deep Inelastic Scattering (DIS) where a high-energy probe interacts with one of the quarks inside a proton, there is a strong $u$-quark dominance so $D_{LL'} \approx D_{LL',u}$ and may be used as a measure of non-strange contributions to the $\Lambda$ polarization \cite{Airapetian_2006}.


\section{$\Lambda$ Signal Selection} \label{Selection}
The CLAS12 spectrometer \cite{Burkert_2020} at Jefferson Lab in Newport News, Virginia operates in tandem with Jefferson Lab’s recently upgraded Continuous Electron Beam Accelerator Facility (CEBAF) \cite{cebaf} to study a broad range of topics in high-energy electron nucleon scattering.  The CLAS12 detector provides a wide polar angle coverage and full azimuthal coverage and has excellent momentum resolution and Particle Identification (PID) capabilities for both charged and neutral particles \cite{Burkert_2020}.  The detector contains a large superconducting toroidal magnet for momentum determination which may be operated in one of two different configurations: inbending (negative particles bent toward the beamline) or outbending (negative particles bent away from the beamline).  CEBAF provides a high-energy polarized electron beam, which interacts with a proton or deuteron target, and final state particles may be reconstructed with the information from the detector subsystems.

Events were selected based on the detection of a proton, a pion, and an electron.  All pairwise combinations of identified protons and pions in a given event were considered.  The proton carries most of the $\Lambda$ momentum, so in the inbending toroidal field configuration the low-momentum, negatively charged pions are lost in the beam line.  Hence, the $\Lambda$ signal is highly suppressed for the inbending configuration and only outbending data were used for this analysis.

\begin{figure}[tbh]
\centering
\includegraphics[width=0.35\textwidth]{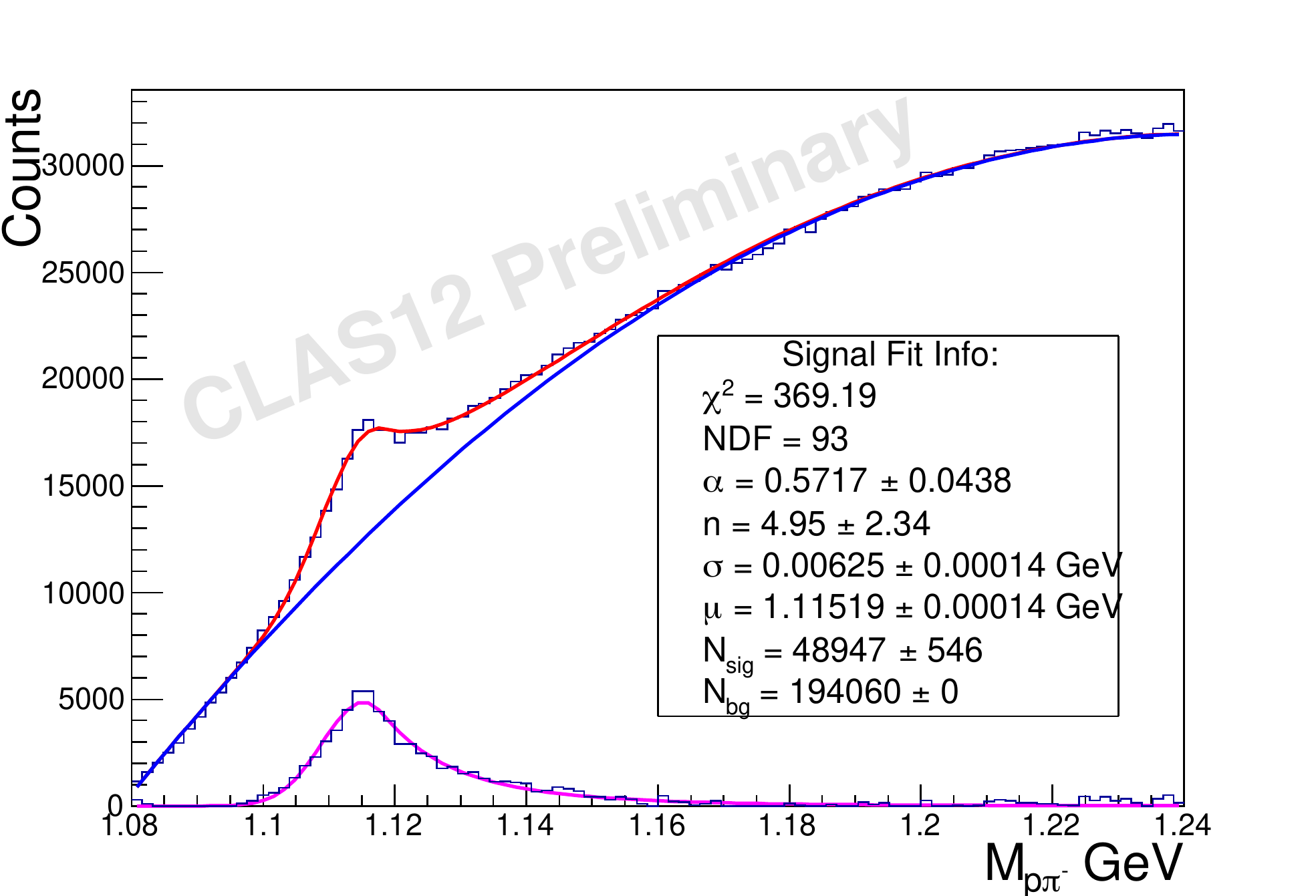}
\includegraphics[width=0.35\textwidth]{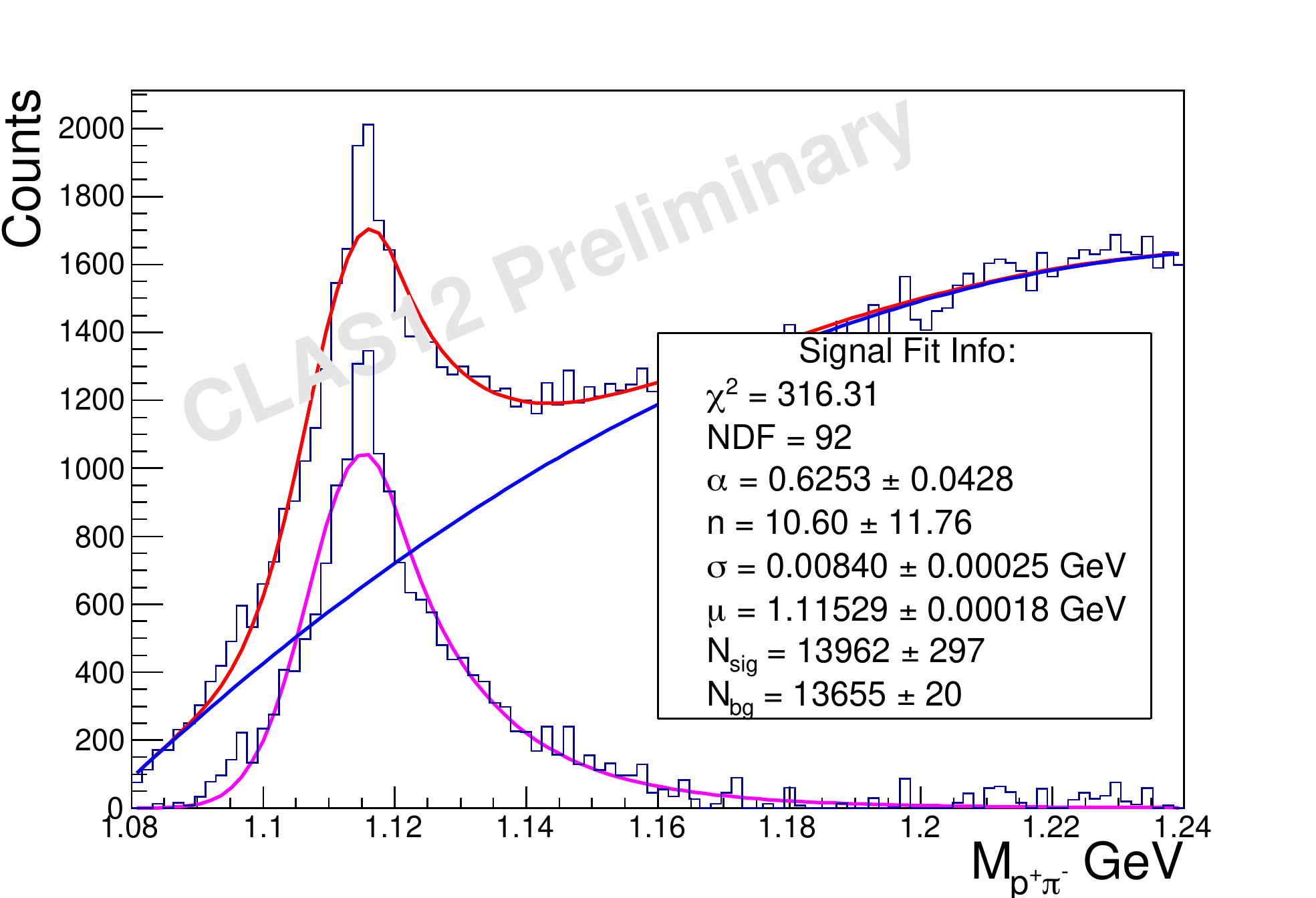}
\caption{A Crystal Ball fit over a quadratic background (red line, blue line is the fitted background) was used to extract the $\Lambda$ signal (magenta line) from data (left) and MC (right).}
\label{fig:MassFit}
\end{figure}

The $p\pi^-$ invariant mass distribution is shown in Fig. \ref{fig:MassFit} for a roughly $\text{20}\%$ subset of available data, collected with a $10.6$ GeV polarized electron beam, unpolarized hydrogen target, and an outbending toroidal field.  Also shown in Fig. \ref{fig:MassFit} is the corresponding fit on simulated Monte Carlo (MC) events.  Standard kinematics selection criteria (momentum transfer $Q^2>1$ GeV$^2$, hadronic final state mass $W>2$ GeV, fractional energy of the virtual photon $y<0.8$, and fractional energy of the $\Lambda$ $z<1$) were applied.  To select events from the current fragmentation region, the requirement Feynman $x_F>0$ was imposed  (where $x_F \equiv p_{||}/p_{|| \, max}$ is the ratio of the  $\Lambda$ longitudinal momentum $p_{||}$ in the center-of-mass frame to the maximum possible $p_{||}$, in this case that of the virtual photon).

A Crystal Ball fitting function \cite{Gaiser_1982} was used to extract the signal width and background ratios as it modeled the data more closely than a simple Gaussian signal fit.  The $\Lambda$ signal is much more pronounced in MC, yet fit shapes were consistent between MC and data.  To estimate the number of signal events we used the excess counts over the quadratic background rather than the integral of the signal fit.  The signal region was chosen by varying the location of different width intervals about the signal peak to find the one that maximized the figure of merit $N_{sig}/\sqrt{N_{bg}}$.  The identified signal region was a $\pm2\sigma$ interval about the fit mean $M=1.115$ GeV.  Sidebands for background subtraction were chosen to be $1.08\text{--}1.10$ GeV and $1.15\text{--}1.18$ GeV to be sufficiently far from the signal tail.

\section{Graph Neural Networks} \label{GNNs}
Standard methods such as secondary vertex reconstruction have so far not been successful at reducing the background in the invariant mass spectrum of the $\Lambda$ at CLAS12 because the current tracking resolution for the low-momentum pions is too low.  Using Armenteros-Podolanski plots \cite{Armenteros_1954} to select $\Lambda$ events also proved to be problematic because the signal shape became too complex to perform a good fit for background estimation.  Hence, we focused our efforts on GNNs.  Our training sample consisted of 96k events from the MC outbending sample filtered so that $50\%$ were true $\Lambda$ signal events and the other half were true background.  Events were taken in the $\pm2\sigma$ signal region obtained from the fit on the MC mass spectrum, with $75\%$ being used for training and the remaining $25\%$ reserved for model validation.  Events were pre-processed into fully connected graphs containing all the particles from the reconstructed event.  The actual data at each node included the reconstructed transverse momentum, polar and azimuthal angles, and PID.  Continuous quantities such as $p_T$, $\theta$, and $\phi$ first had the event mean for that variable subtracted from them and were then normalized to the maximum difference from the event mean.  Discrete quantities such as the PID were reassigned to float values from $-1$ to $1$.  The renormalization of inputs to small values is a standard step in the machine learning process to avoid large gradient values during optimization.

\begin{figure}[tbh]
\centering
\includegraphics[width=0.40\textwidth]{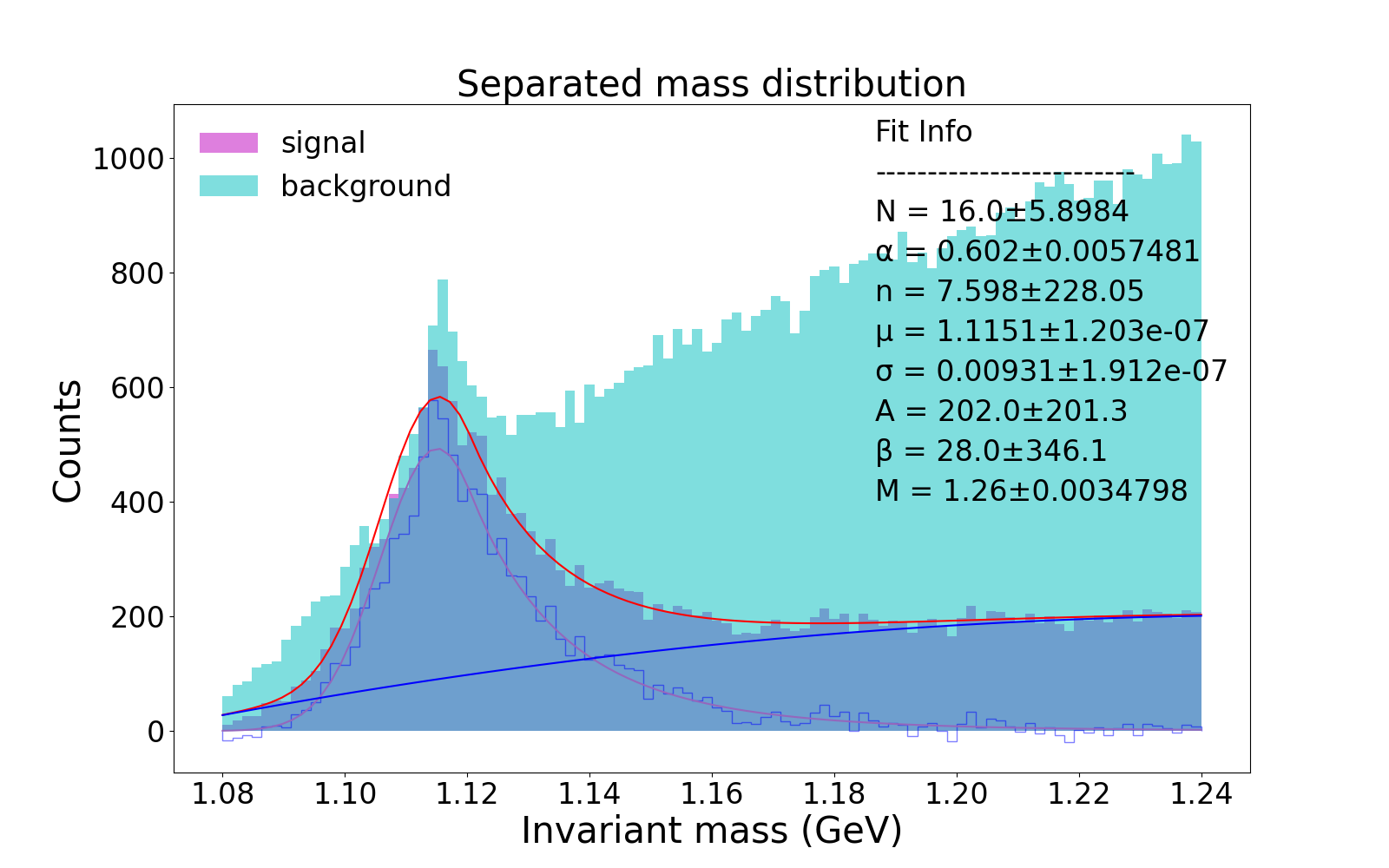}
\caption{We applied the same Crystal Ball signal over quadratic background fit to the GNN-identified signal mass spectrum (dark blue).  The mass spectrum before GNN application (cyan) is shown for comparison.}
\label{fig:GNNFit}
\end{figure}

We first tried the Particle and Energy Flow Network architectures (see Ref. \cite{Komiske_2019}); however, they could not distinguish well between signal and background in training on the MC sample, only reaching test accuracies of $55\text{--}60\%$.  These models were developed with LHC data, which typically see hundreds of particles per event.  In comparison, our dataset has on the order of 10 particles reconstructed per event, so we believe this relative sparsity of the input data is the reason for poor performance on CLAS12 events.  We next turned to Graph Isomorphism Networks (GINs), which are theoretically the most powerful form of GNNs and have been quite successful at graph classification tasks \cite{xu2019powerful}.  After training for up to 100 epochs, the GIN models distinguished between $\Lambda$ signal and background with about 85\% accuracy.  Importantly, the Crystal Ball fit parameters characterizing the signal shape ($\alpha$, $n$, $\sigma$, $\mu$) from the GNN-identified signal were in good agreement with those found without GNN signal selection but with the background reduced by a factor of $4$ as shown in Fig. \ref{fig:GNNFit}.  Note that the results presented later in this paper were \textbf{not} obtained using GNN methods.  Further work is needed to extend GNN performance to actual data.

\section{Determination of $D_{LL'}$} \label{Measurement}
The hyperon polarization can be determined through the angular distribution of the decay protons as:
\begin{equation}
    \frac{dN}{d\Omega_p} \propto 1 + \alpha\vec{P}_{\Lambda} \cdot \hat{n}_p = 1 + \alpha P_b D(y) D_{LL'}^{\Lambda} \cos{\theta_{pL'}}
\label{eq:HBEq}
\end{equation}
where $\alpha=0.732\pm0.014$ is the asymmetry parameter for the weak decay of the $\Lambda$ in its center of mass frame \cite{PDG_2021} and $\vec{P}_{\Lambda} \cdot \hat{n}_p$ is the momentum of the $\Lambda$ along the proton momentum \cite{Airapetian_2006,Guan_2019}.  The variables $P_b$, $D(y)=\frac{1-(1-y)^2}{1+(1-y)^2}$, $\cos{\theta_{pL'}}$ correspond to the event helicity, depolarization factor and the cosine of the angle between the proton momentum and $\Lambda$ spin quantization axis in the $\Lambda$ frame.  Two different choices for the quantization axis are as follows: along the $\Lambda$ momentum or along the virtual photon momentum.  We produce results for both axis choices.

Two different methods were used to calculate the spin transfer coefficient $D_{LL'}$.  The first was an event-by-event method (helicity balance method) detailed in Ref. \cite{Schnell_1999}, which is based on the method of maximum likelihood and the requirement that the luminosity averaged beam polarization $\bar{P}_b$ be consistent with zero.  For a given bin with $N_{\Lambda}$ events, the spin transfer coefficient is calculated as:
\begin{equation}
    D_{LL'} = \frac{1}{\alpha \overline{P_b^2}}\frac{\sum_{i=1}^{N_{\Lambda}}P_b D(y_i) \cos{\theta_{pL'}^i}}{\sum_{i=1}^{N_{\Lambda}}D^2(y_i) \cos^2{\theta_{pL'}^i}}
\label{eq:HBeq}
\end{equation}
where $\overline{P_b^2}$ is the luminosity-averaged squared beam polarization, which was $(89.22\pm2.51)\%$ for our data \cite{Hayward_2021}.  The advantage of this method is that no acceptance correction is needed.

The spin transfer coefficient may also be determined by a simple linear fit on the $\cos{\theta_{pL'}}$ distribution. However, this method requires an acceptance correction for possible false asymmetries, which can be achieved by dividing the $\cos{\theta_{pL'}}$ distribution of one helicity by the other assuming the acceptance does not depend on helicity.  Histogram bin uncertainties were calculated assuming a Poissonian variance $1/\sqrt{N_\pm}$ for the counts $N_\pm$ in a given bin for a single helicity.  From the variance for a binomial distribution, the uncertainty for the corresponding acceptance-corrected bin would thus be $(N_+/N_-)\,\sqrt{1/N_++1/N_-}$.  The spin transfer $D_{LL'}$ can then be calculated from the fitted slope $B$ and constant offset $C$ of the distribution in $\cos{\theta_{pL'}}$ as $D_{LL'} = B/(\alpha \, C)$.

The spin transfer was calculated with these two methods for both choices of spin quantization axis for events within a $\pm2\sigma=\pm0.0125$ GeV window about the signal mean $\mu=1.115$ GeV.  To isolate the $\Lambda$ contribution from background, results for each kinematic bin were corrected as:
\begin{equation}
    D_{LL'}^{\Lambda} = \frac{D_{LL'} - \epsilon D_{LL'bg}}{1 - \epsilon}
\label{eq:BGeq}
\end{equation}
where $D_{LL'bg}$ is the background contribution collectively from both sidebands and $\epsilon = \frac{N_{bg}}{N_{\Lambda}+N_{bg}} = 0.80$ is the fraction of background events estimated from the Crystal ball fit.


\section{Systematic Uncertainty Estimations} \label{Systematics}
To assess the systematic uncertainty in our results, the spin transfer coefficient $D_{LL'}$ was calculated for all oppositely charged pairs assuming they were a $p\pi^-$ pair without restricting particle ID.  The resulting spin transfer was found to be consistent with zero in the signal region, and in the sideband regions, the spin transfer was consistent with zero both with and without particle ID requirements.  The effect of the fit uncertainty on the results was also assessed by varying the fraction of background contribution $\epsilon$ obtained from the fit, but was found to be on the order of $<0.001$.

To compare the results from the helicity balance and linear fit methods the pull between the results was calculated for each kinematic bin as the difference in spin transfer coefficients normalized to systematic uncertainty. If the results from the two methods have respective uncertainties $\sigma_A$ and $\sigma_B$ and the difference in results is $\Delta D_{LL'}$ then the pull is $\Delta D_{LL'}/\sqrt{|\sigma_A^2-\sigma_B^2|}$ assuming the two sets of results are completely correlated.  In general, results in each kinematic bin were within $\pm2\sigma$ of statistical uncertainty between the two methods and clearly show consistent trends so it is assumed that using the fully correlated uncertainty is a valid approach.  However, since the measures of systematic uncertainty considered were all significantly less than the statistical uncertainties in each kinematic bin, systematic uncertainties were not included in our results shown here.


\section{Results} \label{Results}
\begin{figure}[tbh]
\centering
\includegraphics[width=0.43\textwidth]{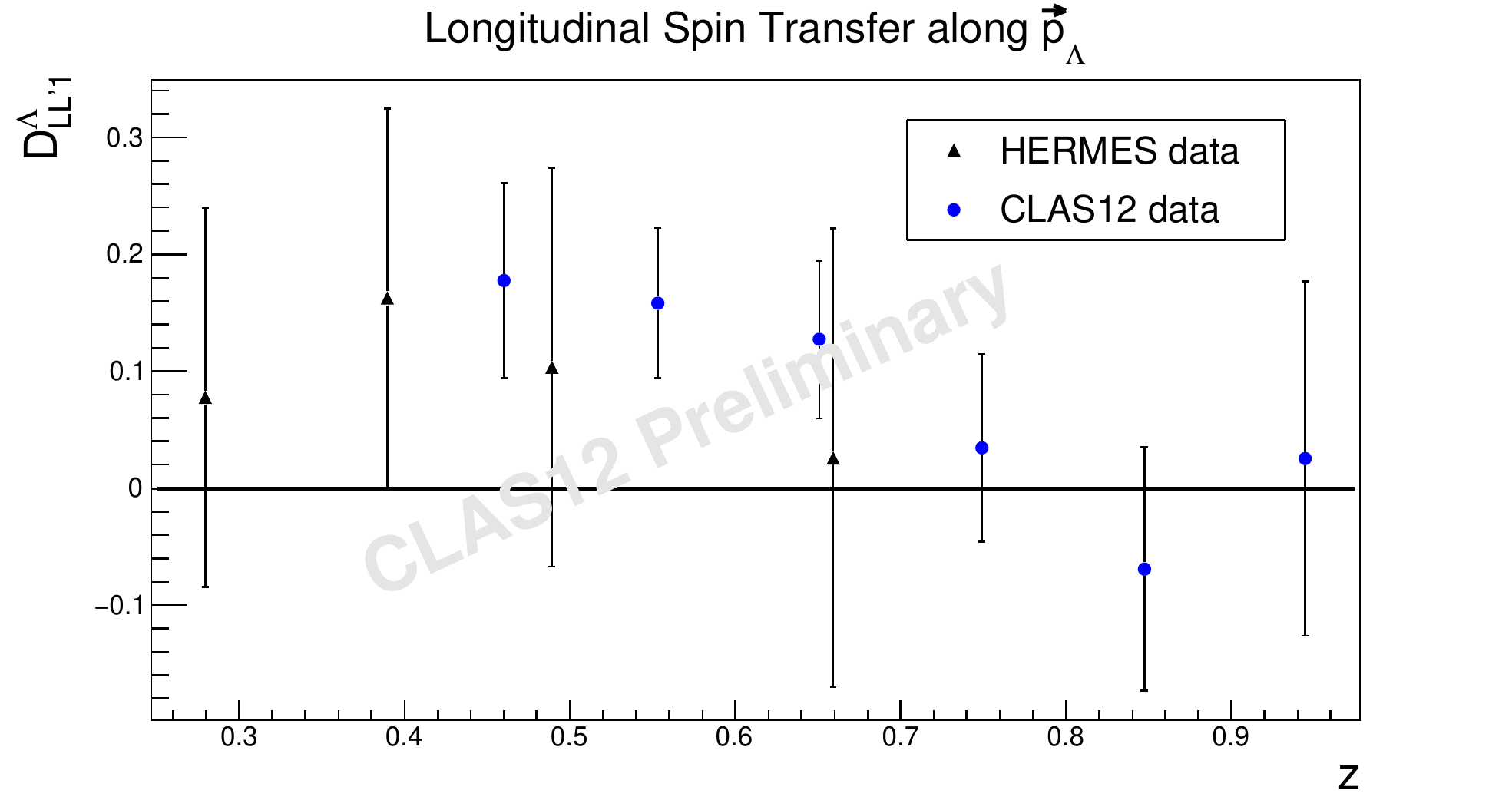} 
\includegraphics[width=0.43\textwidth]{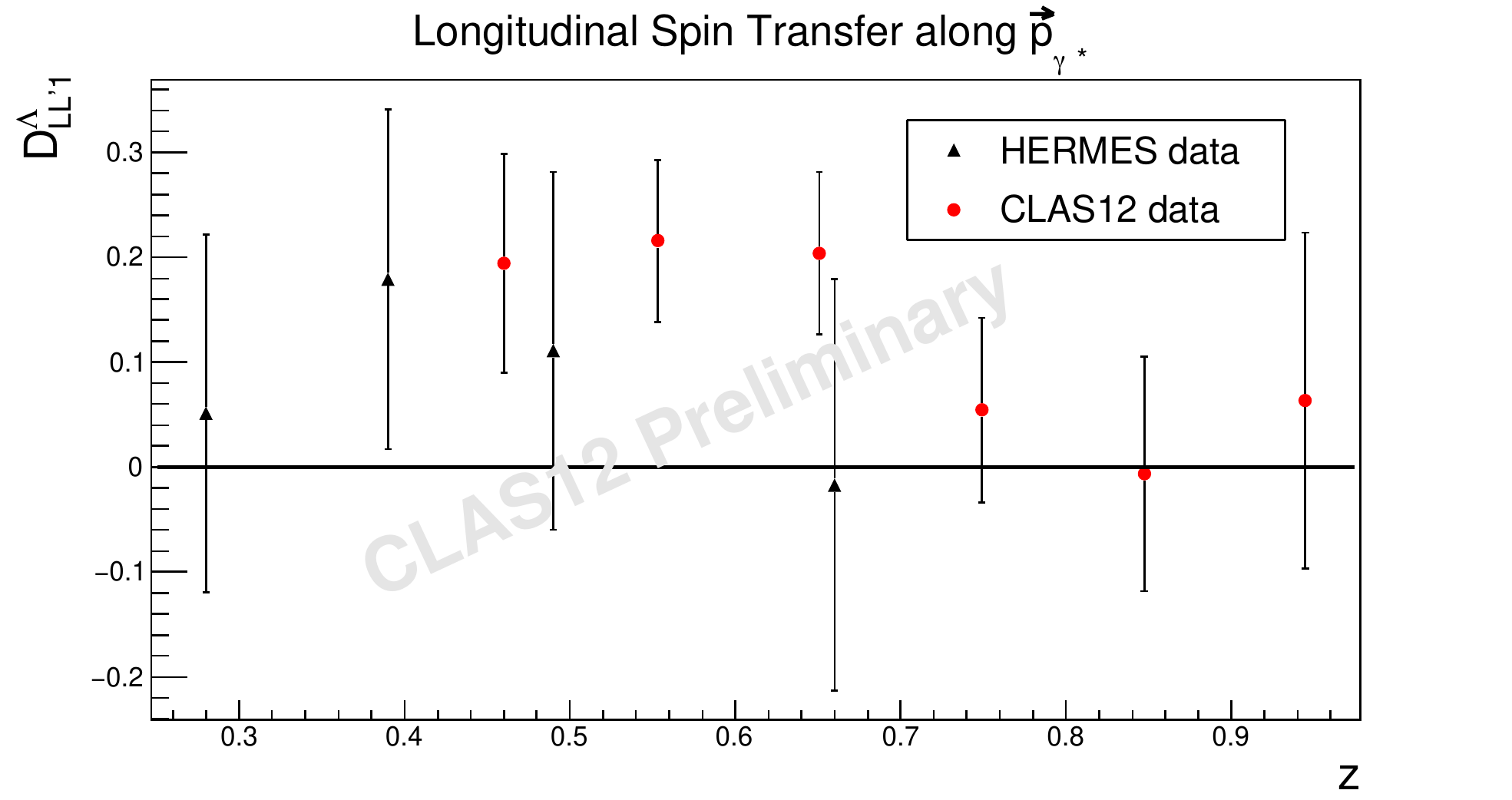} 
\caption{$D_{LL'}^{\Lambda}$ measurements from the HERMES experiment \cite{Airapetian_2006} binned vs. $z$ are in generally in agreement with helicity balance results from CLAS12 for $\cos{\theta_{pL'}}$ along $\vec{p}_{\Lambda}$ (left) and $\vec{p}_{\gamma^*}$ (right).  Note HERMES results were computed with an outdated value for the asymmetry parameter $\alpha=0.624\pm0.013$ and have been rescaled.}
\label{fig:hbz}
\end{figure}

$D_{LL'}$ results obtained using the helicity balance method are shown binned in $z$ in Fig. \ref{fig:hbz}.  The kinematically averaged results for the helicity balance method were $D_{LL'}=0.0618 \pm 0.0963$ $ (\text{stat})$ for $\cos{\theta_{pL'}}$ along $\vec{p}_{\Lambda}$ and $D_{LL'}=0.118 \pm 0.107$ $(\text{stat})$ for $\cos{\theta_{pL'}}$ along $\vec{p}_{\gamma^*}$.  These measurements are consistent with the findings from the HERMES and NOMAD experiments, which measured kinematically averaged spin transfer coefficients $D_{LL'}^{\Lambda}0.11 \pm 0.10$ $ (\text{stat}) \pm 0.03$ $(\text{syst})$ in $e^-$ DIS \cite{Airapetian_2006} and $P^{\,\mu}_{\Lambda}=-0.09 \pm 0.06$ $(\text{stat}) \pm 0.03$ $(\text{syst})$ in $\nu_{\mu}$ DIS (for $x_F>0$) \cite{Naumov_2001}, respectively.  Although data from these other experiments tend to be in a slightly different region of phase space, particularly lower $z$ values, the trends of our data in the overlapping regions are consistent with these previous results and drastically improve on their statistical precision.

\section*{Acknowledgements}
We acknowledge the outstanding efforts of the staff of the Accelerator
and the Physics Divisions at Jefferson Lab in making this experiment
possible.

\paragraph{Funding information}
This work was supported in part by the U.S. Department of Energy, the
National Science Foundation (NSF), the Italian Istituto Nazionale di
Fisica Nucleare (INFN), the French Centre National de la Recherche
Scientifique (CNRS), the French Commissariat pour l$^{\prime}$Energie
Atomique, the UK Science and Technology Facilities Council, the National
Research Foundation (NRF) of Korea, the Helmholtz-Forschungsakademie
Hessen für FAIR (HFHF) and the Ministry of Science and Higher Education
of the Russian Federation. The Southeastern Universities Research
Association (SURA) operates the Thomas Jefferson National Accelerator
Facility for the U.S. Department of Energy under Contract No.
DE-AC05-06OR23177.
The work of MM is supported by the U.S. Department of Energy, Office of
Science, Office of Nuclear Physics under Award Number DE-SC0019230.


\end{document}